\documentclass[12pt,mathptm,times]{article}
\usepackage{graphicx}
\usepackage{setspace}
\usepackage[numbers]{natbib}
\usepackage{amsmath, amsthm, amsfonts, amssymb}

\usepackage{soul} 

\usepackage{authblk}
\usepackage{amsmath}
\usepackage{amssymb}
\usepackage{graphicx}
\usepackage{mathrsfs}
\usepackage{enumerate}
\usepackage{pslatex}\usepackage{bm}
\usepackage{color}

\global\long\def\bi{\beta_{0}}
\global\long\def\ti{\tau_{I0}}
\global\long\def\tic{\tau_{I0}^{c}}

\title{Limited role of spatial self-structuring in emergent trade-offs during pathogen evolution}
\author[1,2]{V\'{\i}ctor Buend\'{\i}a}
\author[1]{Miguel A. Mu\~noz}
\author[3,4,*]{Susanna Manrubia}
\affil[1]{Departamento de Electromagnetismo y F\'{\i}sica de la Materia e Instituto Carlos I de
  F\'{\i}sica Te\'orica y Computacional, Universidad de Granada, E-18071 Granada, Spain}
\affil[2]{Dipartimento di Scienze Matematiche, Fisiche, e Informatiche, Universit\`a di Parma, via G.P. Usberti, 7/A - 43124, Parma, Italy}
\affil[3]{Grupo Interdisciplinar de Sistemas Complejos (GISC), Madrid}
\affil[4]{Programa de Biolog\'{\i}a de Sistemas, Centro Nacional de Biotecnolog\'{\i}a (CSIC),
  Madrid, Spain}
\affil[*]{Corresponding author}

\usepackage{geometry}
\begin{document}

\maketitle

\doublespacing

\begin{abstract}
  Pathogen transmission and virulence are main evolutionary variables broadly assumed to be linked through trade-offs. In well-mixed populations, these trade-offs are often ascribed to physiological restrictions, while populations with spatial self-structuring might evolve emergent trade-offs. Here, we reexamine a model of the latter kind proposed by Ballegooijen and Boerlijst with the aim of characterising the mechanisms causing the emergence of the trade-off and its structural robustness. Using invadability criteria, we establish the conditions under which an evolutionary feedback between transmission and virulence mediated by pattern formation can poise the system to a critical boundary separating a disordered state (without emergent trade-off) from a self-structured phase (where the trade-off emerges), and analytically calculate the functional shape of the boundary in a certain approximation. Beyond evolutionary parameters, the success of an invasion depends on the size and spatial structure of the invading and invaded populations. Spatial self-structuring is often destroyed when hosts are mobile, changing the evolutionary dynamics to those of a well-mixed population. In a metapopulation scenario, the systematic extinction of the pathogen in the disordered phase may counteract the disruptive effect of host mobility, favour pattern formation and therefore recover the emergent trade-off.
\end{abstract}


\section{Introduction}

Host-parasite systems are widespread in nature. Understanding the selection mechanisms underlying virulence and pathogen transmissibility is essential to control outbreaks and disease progression in the long term~\cite{messinger_consequences_2009}. Humans, livestocks, and crops are the most studied hosts due to their relevance for human activities. Despite some commonalities, different hosts differ profoundly in strategies that have coevolved with pathogens, such as immunity or avoiding behaviour, but also in intrinsic features, such as their degree of mobility. As evidence supports, pathogens specific of different host categories display diverse adaptive strategies~\cite{dodds:2010,brunke:2014,lucia-sanz:2017}. 

Early models in epidemiology were developed two centuries ago, at a time when empirical data was scarce~\cite{hethcote:2000}. A modelling standard was much later set by compartmental epidemiological models~\cite{anderson_coevolution_1982}, which classify the individuals in a population in a few states (susceptible, infected, recovered) and are usually formulated as a set of differential equations at the mean-field level --i.e. assuming a well-mixed population. In the last forty years, many efforts have been devoted to devise and analyse mathematical and computational models able to explain and predict the evolution of viral traits~\cite{ewald_host-parasite_1983,anderson_coevolution_1982,alizon_transmissionrecovery_2008,kamo_role_2007}. Recently, the ability to obtain and process big data, along with the development of detailed metapopulation models, has allowed for the achievement of reasonably accurate estimations of epidemic spreading in the short term~\cite{tizzoni_real-time_2012,zhang_spread_2017}.

Pathogens are in need of suitable strategies in order to persist in a host population, in a continuous arms race with the defenses of the host. Understanding which strategies are selected along evolution may help developing improved epidemiological models that include the long-term behaviour of the strains. The classical theory~\cite{alizon_transmissionrecovery_2008,lion_evolution_2015} assumed that new pathogens, not adapted to the host, are more virulent (where virulence is understood as the number of deaths caused by the disease). Indeed, mean-field epidemiological models predict that the base reproductive number $R_{0}$, defined as the average number of infections produced by an infected individual, tends to increase along evolution~\cite{anderson_coevolution_1982}. Maximization of $R_{0}$ in a finite population, however, exhausts the pool of susceptible individuals, leading to pathogen extinction. A possible way out of this runaway process would be for the pathogen to become less virulent as it co-evolves with its host; however, a revision of experimental data shows that this hypothesis does not hold in general~\cite{ewald_host-parasite_1983,anderson_coevolution_1982}. As an alternative solution, it was proposed that pathogens should be subjected to a trade-off between virulence and transmission, such that virulent strains have a lower transmission rate~\cite{alizon_transmissionrecovery_2008}. The equilibrium between an aggresive, virulent strategy and a low-virulence, highly transmissive strategy results in a limit to the reproduction speed, i.e. a maximum value of the base reproductive number~\cite{alizon_transmissionrecovery_2008,kamo_role_2007}, yielding a possible explanation for the existence of strains with intermediate values of $R_0$. On the other hand, the trade-off hypothesis does not explain avirulent diseases. For this case, alternative trade-offs have been proposed, such as the recovery-transmission trade-off~\cite{alizon_transmissionrecovery_2008,shrestha_evolution_2014}. The trade-off hypothesis has been questioned in the last years for several reasons. One of the most important issues regards a definition of virulence able to link observed data with theoretical models. As a consequence, as of yet, there is little available empirical evidence supporting the existence of such physiological trade-offs~\cite{alizon_virulence_2009}. 

The transmission of pathogenic diseases rarely occurs in a host population that is well mixed, such that an important issue in infection propagation is the role played by structured host populations~\cite{lion:2008} and the evolutionary parameters characterizing pathogen strategies~\cite{sun_pattern_2016}. Many models have studied the effect of space and network structure in disease spreading~\cite{watts_multiscale_2005, keeling_implications_2005}. There is ample theoretical evidence that the evolutionarily stable traits of pathogens significantly differ in well-mixed or spatially structured scenarios~\cite{messinger_consequences_2009} and that space has important effects on evolutionary dynamics~\cite{boots:2004,webb:2007,boerlijst:2010} as well as on the effects of different functional forms of the {\it a priori} trade-off between transmission and virulence~\cite{webb:2013}.

Several studies suggest that trade-offs do not necessarily come from the physiology of the pathogens, but might arise as an emergent property of a spatially structured host-pathogen system~\cite{haraguchi:2000,rauch:2002,ballegooijen_emergent_2004,kamo_role_2007,lion_evolution_2015}. The phenomenology of spatially (self-)structured host-pathogen systems has important features that sets them apart from mean-field approaches. The exhaustion of hosts by highly pathogenic variants becomes localized both in space and in time; through evolution, a hierarchy of time scales related to mutants of decreasing virulence defines a collective and time-dependent invasion fitness~\cite{rauch:2002}. Also, it has been found that the spatial structure is enough to select for intermediate virulence values~\cite{kamo_role_2007}, and that the recovery-transmission trade-off can emerge due to spatial structuring of the hosts in a lattice, even when the infection characteristic time and the transmission rate evolve independently~\cite{ballegooijen_emergent_2004}.

The effects of spatial structure on the evolutionary parameters of pathogens and the possible existence of emergent trade-offs has recently received empirical support.  Experiments with two variants of the phage $\lambda$ infecting {\it E. coli} have demonstrated that spatial structuring impedes the spread of the virulent variant, and selects for a prudent infection strategy~\cite{berngruber:2015}. Also, it has been shown that high host availability favours parasites with lower virulence and higher transmissibility, while low host availability selects for the contrary relationship, experimentally demonstrating the existence of a trade-off promoted by spatial structure~\cite{leggett:2017}. 

Though a few studies have emphasized the important role played by host-mobility fluxes under a metapopulation structure~\cite{poletto_characterising_2015,poletto_host_2013}, host mobility is not typically included in evolutionary models. Interestingly, the effects of host mobility on pathogen infectivity have been empirically evaluated in a series of experiments where insect larvae living in environments that limited their mobility to different extents were infected with a species-specific virus~\cite{boots:2007}: in agreement with expectations, infectivity was reduced concomitantly with host mobility. 

Here, we address the question of the limits to the emergence and the structural stability of the trade-off between transmissibility and virulence under a variety of scenarios that represent realistic features of different host or pathogen types: local diffusion and long-range jumps of hosts (as for cattle or foraging animals, e.g.), high mutation rates (as for virus or viroids), and metapopulation structure. To this end we implement the model by Ballegooijen and Boerlijst~\cite{ballegooijen_emergent_2004} and first characterize, in terms of invasibility criteria, the feedback evolutionary mechanism that selects for a curve of constant $R_{0}$. We show that this curve is a critical self-evolved boundary~\cite{rand_invasion_1995}, whose functional form we calculate in a one-dimensional approximation, that separates two regions with qualitatively different spatial structure. While that evolutionary outcome is mostly robust under variations in the parameters, it is very fragile under host mobility. Finally, we extend our main results to a metapopulation model, showing that the main mechanism of evolution in the metapopulation is infection propagation within each subpopulation. Self-structured and disordered (mean-field) states are characterized by different average lifetimes that entail an asymmetric invasion likelihood such that, in metapopulations with mobile hosts, convergence to the critical boundary is favoured by the metapopulation structure. 

\section{Model}

\begin{figure}[h]
\begin{center}
\includegraphics[width=0.6\textwidth]{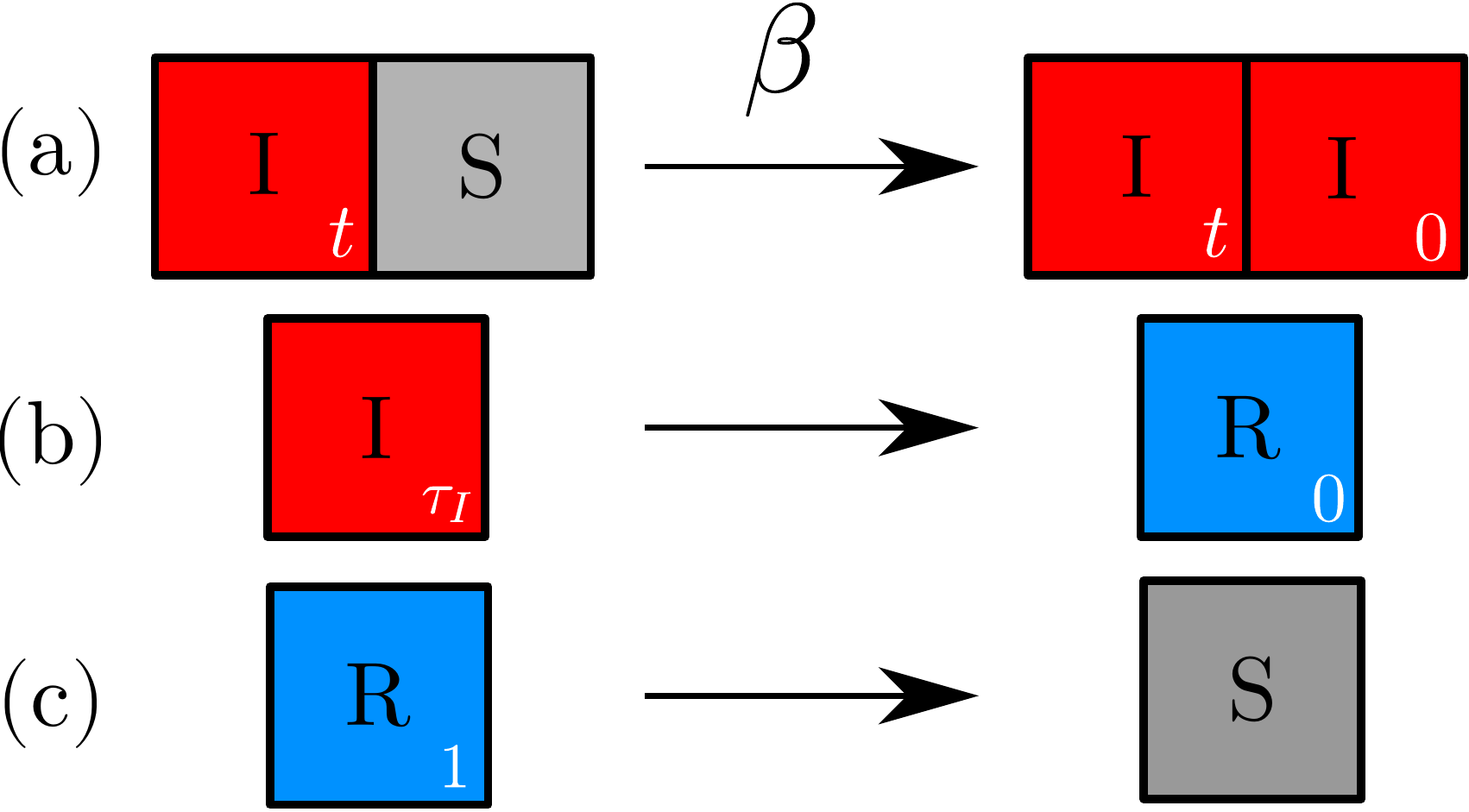} 
\end{center}
\caption{\label{fig:model}Schematic description of the model. Each site in the lattice can be in one of the three states, infected ($I$, red), susceptible ($S$, gray) or recovered ($R$, blue). The colour code for states is maintained all through this work. Each infected site corresponds to an infected host characterized by a pair of values $s=\left(\beta, \tau_I \right)$. An internal time counter for each site triggers reactions (b) and (c). The time counter is reset to zero when the state changes.}
\end{figure}

Following~\cite{ballegooijen_emergent_2004}, we define a host-pathogen dynamical model on a two-dimensional lattice with Moore neighbourhood (each node has $8$ neighbours) and periodic boundary conditions. Each individual occupies a lattice site that can be either in a susceptible ($S$), infected ($I$) or recovered ($R$) state; transitions between these states are controlled by the stochastic reactions

\begin{subequations}
\begin{align}
S+I\overset{\beta}{\rightarrow}2I \label{eq:model1} \\
I\overset{\tau_{I}}{\rightarrow}R \label{eq:model2} \\
R\overset{\tau_{R}}{\rightarrow}S \label{eq:model3} \, ,
\end{align}
\end{subequations}
where $\beta$ is the rate of infection, and $\tau_{I}$ and $\tau_{R}$ are, respectively, the infection and recovery times (see Fig.~\ref{fig:model}). Each node has an internal time counter in order to trigger reactions (\ref{eq:model2}) and (\ref{eq:model3}). In the well-mixed (mean-field) scenario reactions (\ref{eq:model1})-(\ref{eq:model3}) represent an SIRS model with delay. As in~\cite{ballegooijen_emergent_2004}, we employ a fixed time-step algorithm with synchronous update for numerical simulations. At each time step, we loop over the nodes. If a node is in the susceptible state, we identify its infected neighbours, compute the total infection rate, and infect it with the corresponding probability. If a node is in the infected or recovered states, we check if the internal counter is greater than the infection or recovery times, respectively, in order to change its state. The detailed procedure is explained in Methods.

Each infected node $j$ has its own transmission rate $\beta_{j}$, as well as its own infection period $\tau_{Ij}$. A strain is defined by a pair $s=\left(\beta,\tau_{I}\right)$ with no {\it a priori} imposed trade-off. Both parameters are independently mutated at each time step with probability $\mu\Delta t$, where $\mu$ is the mutation rate and $\Delta t$ is the timestep. We assume superinfection exclusion, so a host individual in state $I$ cannot be infected by a second strain.

\begin{figure}[h]
\begin{center}
\includegraphics[width=0.6\textwidth]{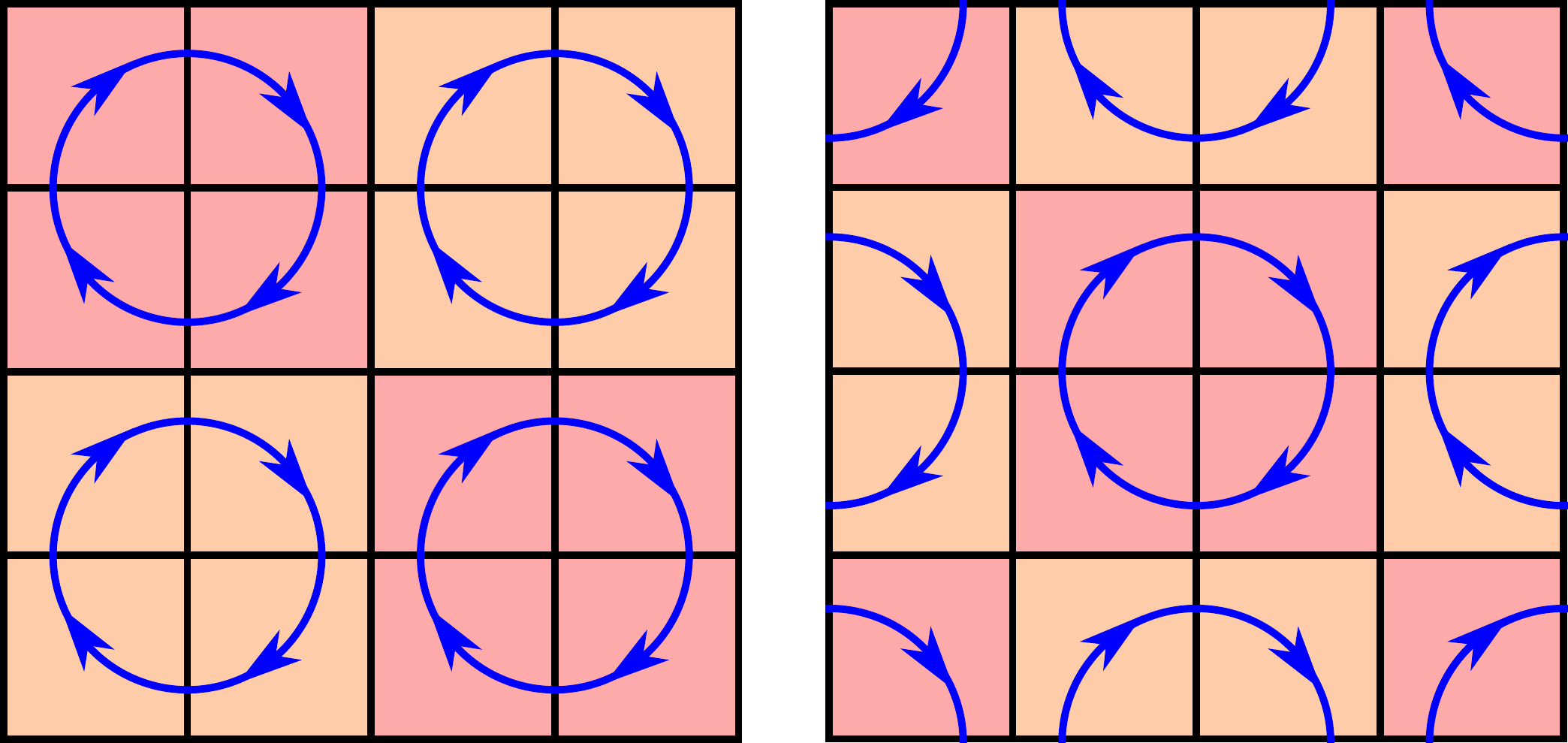} 
\end{center}
\caption{\label{fig:toffoli} Schematic representation of the Toffoli-Margolus algorithm. The leftplot shows how rotation is applied when starting from (0,0); the right plot shows rotation from (1,1). When both steps are alternated, the system is diffusively mixed.}
\end{figure}

To introduce host local diffusion we employ the Toffoli-Margolus algorithm~\cite{toffoli_cellular_1987}. This algorithm divides the lattice in $2\times2$ squares that rotate either clockwise or counterclockwise. Briefly, squares are taken starting alternatively from $(0,0)$ and $(1,1)$ so the system becomes mixed, as illustrated in Figure~\ref{fig:toffoli}, regardless the state of the site.

In a lattice, diffusion is defined as $D=\Gamma(\Delta x)^{2}$, where $\Gamma$ is the rate at which particles hop and $\Delta x=1$ the distance between sites. Since every time the mixing algorithm is applied every particle hops to a neighbouring site, $\Gamma$ can be fixed in order to have the desired diffusion coefficient $D$ as $\Gamma=1/(D\Delta t)$.

To recreate a metapopulation scenario, we use a scheme similar to that implemented in previous studies~\cite{poletto_characterising_2015,poletto_host_2013}. We create a network of lattices of small sizes and connect them in a metapopulation network. After updating all lattices following the previous algorithm, we check the state of each population. Each host (for which the state of a site acts as a proxy) can jump with probability $\lambda\Delta t/N$, where $\lambda$ is the jump rate. If the jump occurs, we randomly select a site in a neighbouring lattice, as shown in Figure~\ref{fig:metapop}. Then, both sites (their states) are exchanged. We fix $\lambda=0.01$. 

\section{Results}

\subsection{Spatial self-structuring results in an emergent trade-off between transmissibility and infectivity}

We start by simulating infection propagation and evolution without host mobility, i.e. with $D=0$, to recall the core results of Ballegooijen and Boerlijst's model~\cite{ballegooijen_emergent_2004}. This delayed SIRS model has base reproductive number $R_{0}=8\beta\tau_{I}$, independent of $\tau_R$, in the mean-field approximation. For fixed parameter values, the system exhibits different kinds of patterns, as depicted in Figure~\ref{fig:patterns}a-d. Low values of $R_{0}$ produce short-lived local infection bursts, while high values of $R_{0}$ eventually lead to the formation of spiral waves, a pattern characteristic of two-dimensional excitable media~\cite{cross_pattern_2009}. Thus, there are two different ``phases'', with wave patterns and without wave patterns, that we call ordered or self-structured and disordered or mean-field, respectively. A ``phase transition'' or ``critical boundary'' separates these two regimes.

When the system is free to evolve through changes in virulence and transmissibility parameters (keeping the recovering time $\tau_R$ fixed), it converges to trajectories of constant $R_{0}^{ev}=\left(6.623\pm0.003\right)$ that, after a transient period of variable duration, become independent of the initial conditions.

The evolutionary trajectory of the system is illustrated in Figure~\ref{fig:patterns}e. This result is in agreement with~\cite{ballegooijen_emergent_2004}, where it was shown that the trade-off is a by-product of the spatial self-structuring of the system, and where the quantity that seems to be under positive selection is the frequency of emission of infection waves, and not $R_0$. Though not explicitly mentioned in~\cite{ballegooijen_emergent_2004}, where the maximum allowed value of the infectivity was $\beta=4$, the evolutionary trajectory continues indefinitely to higher values of $\beta$ and lower values of $\tau_I$ at an increasingly slower but non-arresting pace.

\subsection{Selection for higher frequency of emission of infection waves only succeeds at the local scale}

In order to delve into the feedback mechanism that drives the evolutionary process towards a fixed base reproductive number, we undertake two simulation experiments to evaluate the ability to invade of different strains once the system is spatially organized. No parameter evolution is considered here, since our aim is to quantify to which extent spatial pattern formation enhances or hinders the invasion of analogous populations. To this end we select a focal population with parameters corresponding to each of the four situations represented in Figure~\ref{fig:patterns}a-d. Let us call this strain $s_{0}\equiv\left(\beta,\tau_{I}\right)$. Then, we also select four strains that are nearby in an evolutionary sense. That is, if the evolutionary process would be on, these strains would be one mutational step away from fhe focal population; their parameters are $s_{++}\equiv\left(\beta^{+},\tau_{I}^{+}\right)$, $s_{+-}\equiv\left(\beta^{+},\tau_{I}^{-}\right)$, $s_{-+}\equiv\left(\beta^{-},\tau_{I}^{+}\right)$, and $s_{--}\equiv\left(\beta^{-},\tau_{I}^{-}\right)$, where we have defined $\beta^\pm = \beta \pm \Delta \beta$ and $\tau _I ^\pm = \tau _I \pm \Delta \tau _I$ to simplify the notation. All possible competitions between the focal strain $s_0$ and its mutants are assayed, and each competition is performed 10 times.

Note that the five populations (one focal and four nearby mutants) can be ordered with respect to their base reproductive number. In the simulations, where initial conditions ensure that $\beta > \tau_I$ for all times, the ordering is given by

\begin{equation}
  \label{eq:R0}
R_{0}\left(s_{++}\right)>R_{0}\left(s_{-+}\right)>R_{0}\left(s_{0}\right)>R_{0}\left(s_{+-}\right)>R_{0}\left(s_{--}\right) \, ,
\end{equation}
so $s_{++}$ and $s_{-+}$ increase the base reproductive number of the focal population $s_0$, while $s_{+-}$ and $s_{--}$ decrease it. On the other hand,  if $\beta < \tau_I$, the ordering is $R_{0}\left(s_{++}\right)>R_{0}\left(s_{+-}\right)>R_{0}\left(s_{0}\right)>R_{0}\left(s_{-+}\right)>R_{0}\left(s_{--}\right)$.
In a mean-field scenario, this ordering coincides in either case with the relative advantage of one population over the others and, thus, it determines the mutual ability to invade and predicts the outcome of competition experiments in a well-mixed scenario.

However, results in~\cite{ballegooijen_emergent_2004} suggest that, in cases where the population is spatially structured, the relative advantage corresponds not to the population with the larger $R_0$, but to the one with the higher frequency $w$ of emission of infective waves. The emission frequency is obviously larger for those strains with a faster infection rate which simultaneously produce waves with narrower fronts, but these two quantities do not bear a straight relationship with parameters $\tau_I$ and $\beta$.

In order to better understand how the invasibility criteria might change in spatially structured systems, let us briefly explore the one-dimensional version of model~\cite{ballegooijen_emergent_2004}. The frequency $w_{1D}$ is the inverse of the average time elapsed between two consecutive infection events. A node with an infected neighbour becomes infected after a typical time $t_{\beta}=\beta^{-1}$, and stays itself infected for a time $\tau_I$. Then it recovers to become susceptible again after a time $\tau_R$. Therefore,

\begin{equation}
  \label{eq:w}
  w_{1D}=\frac{1}{1/\beta+ \tau_I+\tau_R} \, ,
\end{equation}
and the mutants and the focal population display the following ordering

\begin{equation}
  \label{eq:worder}
  w_{1D}\left(s_{+-}\right)>w_{1D}\left(s_{--}\right)>w_{1D}\left(s_{0}\right)>w_{1D}\left(s_{++}\right)>w_{1D}\left(s_{-+}\right) \, ,
\end{equation}
which yields invadability criteria different from Eq.~(\ref{eq:R0}). Again, as for the basic reproductive number, the precise ordering of the mutants $s_{--}$ and $s_{++}$ depends on the relative values of the parameters. The order in eq.~(\ref{eq:worder}) corresponds to the case studied in simulations, $\beta>\tau_I$. On the other hand, if $\beta<\tau_I$,  then $w_{1D}\left(s_{+-}\right)>w_{1D}\left(s_{++}\right)>w_{1D}\left(s_{0}\right)>w_{1D}\left(s_{--}\right)>w_{1D}\left(s_{-+}\right)$. This calculation, however, cannot be straightforwardly extended to the two-dimensional case.

\subsubsection{Invadability experiment 1}
To go beyond the one-dimensional case, we considered a two-dimensional lattice of size $2L\times L$; the two parts of the space are not connected initially. $s_0$ and each of its mutant strains are picked up pair-wise and placed either at the left or right half of the lattice. After a fixed time, such that spatial patterns have developed according to the parameters chosen, both parts of the space are allowed to interact.

In all simulations performed under the previous conditions, the system was eventually invaded by the strain with higher $R_{0}$, independently of any other condition. Strains $s_{++}$ and $s_{-+}$ were systematically selected in competition with $s_0$, while strains $s_{+-}$ and $s_{--}$ were always removed from the system, as would be predicted by a mean-field approximation. This is an {\it a priori} unexpected result that apparently contradicts the dynamics of evolutionary trajectories in the spatially structured model. 

\subsubsection{Invadability experiment 2}
Take strain $s_{0}$ and let the system run enough time to develop patterns in a large $L \times L$ lattice. Then, substitute any infected individual inside a randomly chosen area of size $10\times10$ sites with strains of one of the nearby mutants. In this case, we observed two different behaviours:

\begin{enumerate}
\item If the focal strain $s_0$ was not able to develop patterns (that is, it is located in the disordered region of the parameter space, with a base reproductive number below $R_0^{ev}$), the mutant strain invades the whole system if its $R_{0}$ is higher: $s_{++}$ and $s_{-+}$ are at an advantage and therefore are selected in front of $s_0$;
\item If $s_0$ has clearly developed patterns (with a base reproductive number above $R_{0}^{ev}$), the only mutant strain able to invade the system is $s_{+-}$, following the direction observed in the evolutionary curve. 
\end{enumerate}

\subsubsection{The critical boundary emerges as an equilibrium between two different selection mechanisms}

The results above highlight that the likelihood to invade a spatially organized population depends on the size and structure of the invading population. Very often, this size is small because it is a single mutant individual or a small sample of individuals from disconnected populations that attemp the invasion. If this is so, the scenario in our invadability experiment 2, whose dynamical properties coincide with those leading to the emergence of the trade-off, is the applicable one.

\begin{figure}[h]
\begin{center}
\includegraphics[width=1.\textwidth]{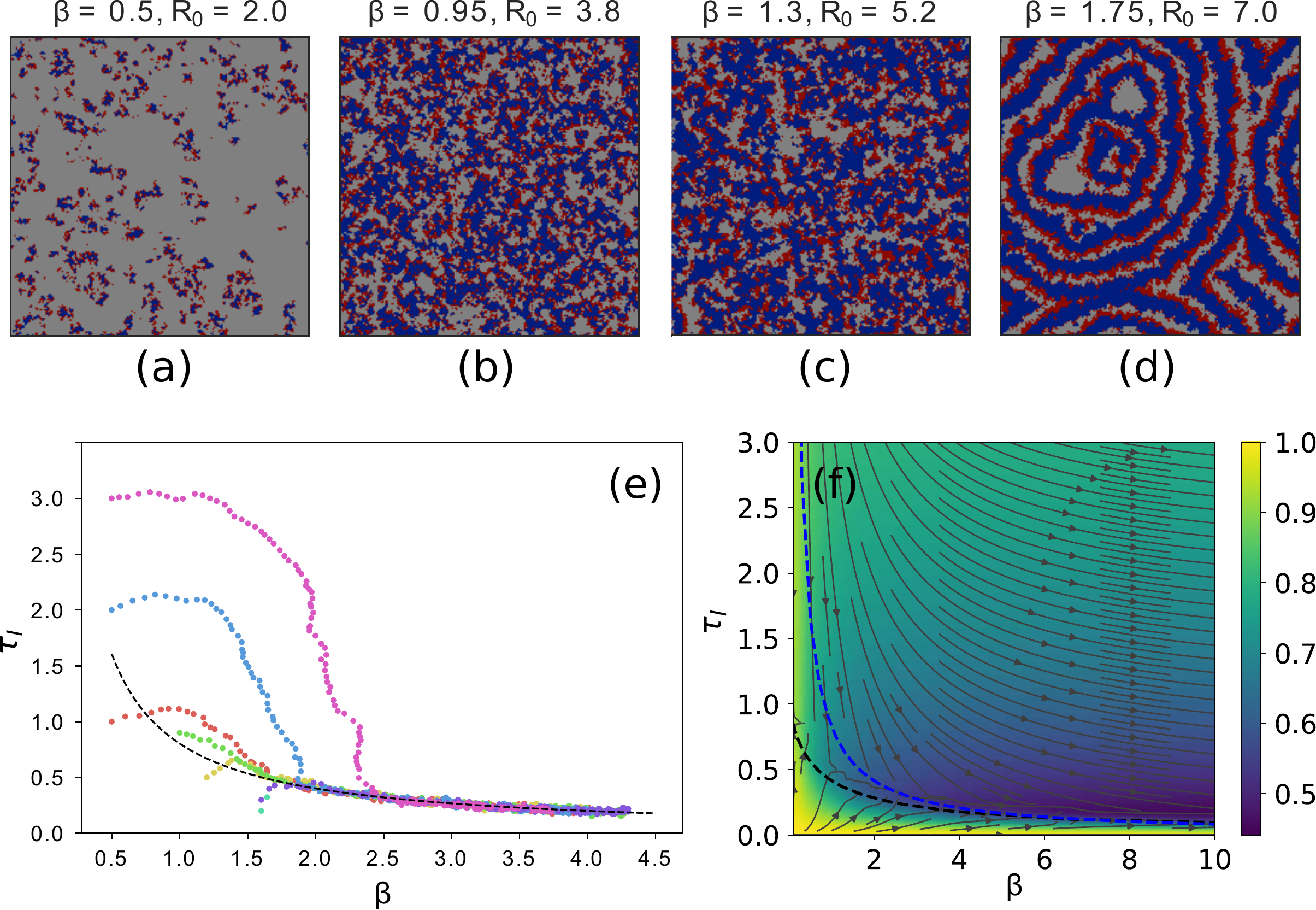} 
\end{center}
\caption{\label{fig:patterns} Pattern formation and evolutionary dynamics in Ballegooijen and Boerlijst's model~\cite{ballegooijen_emergent_2004}. (a-d) Pattern formation as a function of fixed parameter values. In all plots we show a statistically stationary state of a $100 \times 100$ lattice with $\tau_{I}=0.5$ and $\tau_R=1$. For low $R_{0}$ (a,b), only local infections develop. As the value of $R_{0}$ increases, incipient spiral wave patterns appear (c) and become fully developed for larger base reproductive number (d). (e) Evolutionary trajectories for different initial conditions. All trajectories are eventually attracted to the curve $\tau_{I}\left(\beta\right)=R_{0}^{ev}/\left(8\beta\right)$ (dotted black line) with constant base reproductive number $R_{0}^{ev}=\left(6.623\pm0.003\right)$, as determined through a non-linear least-squares fit. Here the lattice size is $150 \times 150$. (f) Evolutionary fluxes in phase space. We have considered the two surfaces $\exp{-R_0}$ and $\exp{-w}$, which function as effective potentials to indicate the directions of evolutionary change; its maximum value at each point, max$\{\exp{-R_0},\exp{-w}\}$ is represented through a colour scale. Arrows indicate the selection gradient. The black dashed line corresponds to the relationship in Eq.~\ref{eq:1dExact}, while the blue dashed line is that observed in the simulations.}
\end{figure}

Let us return to the evolutionary trajectory in Figure~\ref{fig:patterns}e with the previous results in mind. We see that initial conditions with low $R_{0}$ first increase the base reproductive number by selecting mutants mainly in the $s_{++}$ direction, and later converge to the curve evolving along the $s_{+-}$ quadrant. All regions in the evolutionary trajectory where $\tau_I$ increases (and also $R_0$, since $\beta$ always grows) correspond to spatially disordered situations, either because the parameters correspond to the phase with $R_0 < R_0^{ev}$ or because the system is in the initial transient before spatial self-structuring sets in. The increase of $R_0$ progressively drives the system to a new regime where spiral waves start to develop. This qualitative change modifies the criteria for invadability, and selection for waves of higher frequenty $w$ sets in.

At this point, it seems reasonable to assume that the critical boundary results from the equilibration of two mechanisms: selection for larger $R_0$ in the disordered phase and selection for higher $w$ in the ordered phase (see Figure~\ref{fig:patterns}f). Even though the frequency of emission is a complex function of the parameters in two dimensions, and results from 1D cannot be straightforwardly extrapolated, in general, to higher spatial dimensions, let us use the functional forms of $w_{1D} (\beta,\tau_I)$ and $R_0(\beta,\tau_I)$ previously derived to give an estimation of the curve where the two surfaces cross. Note that $R_0$ grows in the $s_{++}$ direction. If our understanding of the evolutionary feedback is correct, this estimation should resemble the numerical boundary $R_0^{ev}$. The relationship between $\beta$ and $\tau_I$ along the critical boundary is defined through $w_{1D}=R_0$,

\begin{equation}
  \label{eq:1dExact}
  n \tau_I \beta = \frac{c}{1/\beta+ \tau_I+\tau_R}  \, 
\end{equation}
where $c$ is a constant required for correct dimensionalization and $n$ is the number of neighbours, which yields

\begin{equation}
\label{eq:expansion}
\beta = \frac{c/n - \tau_I}{\tau_I (\tau_I+\tau_R)} \simeq \frac{c}{n \tau_I \tau_R} + \mathcal{O}\left( \frac{1}{\tau_I ^2} \right) \, .   
\end{equation}

For $\tau_R=1$, to first order in $\tau_I$ we get $n \beta \tau_I \sim c$, where we can identify $n=8$ and $c=R_0^{ev}$. The selection mechanism described puts the system at the edge of wave formation: higher $R_0$ strains tend to be selected in the disordered phase, while higher values of $w$ are selected in the spatially structured phase, such that the curve of constant $R_0$ is where these two mechanisms are at equilibrium (see Figure~\ref{fig:patterns}f). This boundary also represents the limit of validity of mean-field calculations, providing a self-consistent, {\it ad hoc} explanation of why its functional form verifies $R_0 =8 \beta \tau_I$. 

\subsection{Spatial self-structuring is fragile}

The boundary $w_0^{ev}=R_0^{ev}$ separates two phases with qualitatively different spatial structure. Selection mechanisms, as revealed by the invasion criteria in either phase, are different and of opposing sense regarding mutations in $\tau_I$, causing an evolutionary feedback loop that eventually drives the system to a self-evolved phase boundary. The emergence of the trade-off is critically dependent on the development and persistence of spatial self-structuring. Are there conditions under which the latter is not possible, even if evolutionary parameters are in the ordered region? In this section we explore the stability of the emergent trade-off under changes in the mobility of hosts as well as in two model parameters that have been kept constant so far: the mutation rate $\mu$ and the recovery time $\tau_R$.

\subsubsection{Host mobility prevents spatial self-structuring}

Broadly speaking, infection propagation depends on the degree of mobility of hosts: propagation speed, endemicity or optimal evolutionary parameters vary whether hosts are sessile, diffuse locally or perform high-distance jumps. In this section, we explore how locally diffusing hosts and hosts able to jump to arbitrary sites in the lattice affect the formation of spatial structures.  

Diffusing hosts are modeled by means of the Toffoli-Margolus algorithm (see Methods). In the high-diffusion limit, the system becomes well-mixed and the expectation is that it behaves as in mean-field, thus increasing its average $R_{0}$~\cite{anderson_coevolution_1982}. Indeed, simulations with high diffusion follow an evolutionary trajectory along which $\beta$ and $\tau_{I}$ steadily increase. The process continues until all individuals become infected, eventually causing the extinction of the pathogen.

\begin{figure}[h]

\includegraphics[width=1.\textwidth]{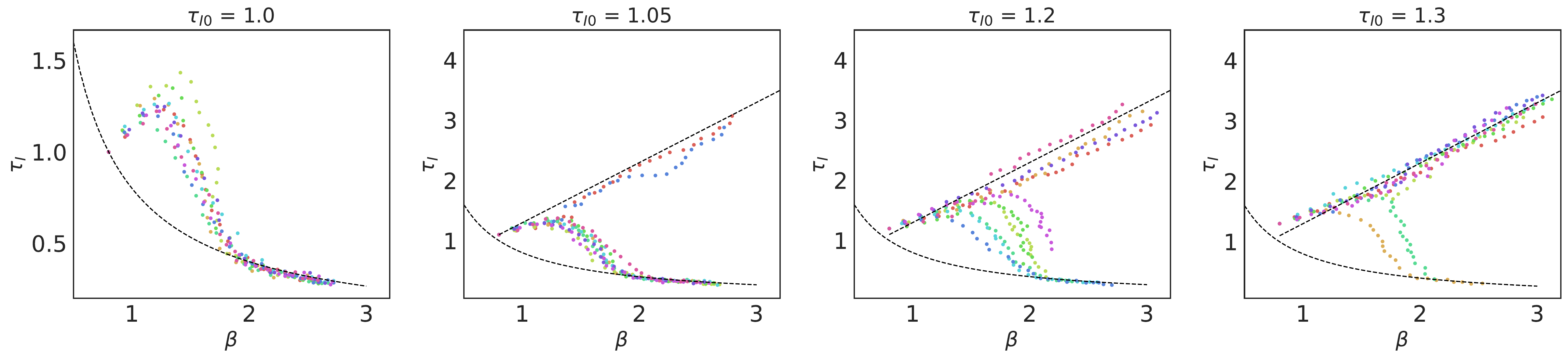} 

\caption{\label{fig:dif} Evolutionary trajectories as a function of the initial infection period. Parameters of the simulations are $D=0.35$, $\beta_{0}=0.8$ and $L=150$, with $\tau_{I0}$ as shown in the legend. The dashed black line shows the expected mean-field behaviour $\tau_I = \beta + \text{constant}$. The system either displays a behaviour indistinguishable from the $D=0$ case or follows a curve of steady increase in $R_{0}$, as predicted in the mean-field theory. The region where stochastic fluctuations can lead the system to any of the two states stretches to a point in the limit $L \to \infty$.}
\end{figure}

On the other hand, sufficiently low diffusion should recover the $D=0$ trajectory we analysed before. Therefore, there should be an intermediate value of the diffusion where the behaviour crosses over from $R_0=R_0^{ev}$ to an ever increasing $R_0$. Our simulations show that, for intermediate $D$ values, the convergence of the system to either phase depends on the initial conditions, as depicted in Figure~\ref{fig:dif}. For an initial fixed value of $\bi$, there exist a critical $\tic$ such that any $\ti<\tic$ will exhibit an evolutionary trajectory identical to $D=0$, while for $\ti>\tic$ the parameters will diverge as predicted by the mean-field theory. The position of the critical point $\tic=\tic\left(\bi\right)$ increases as $\bi$ increases. The persistent perturbation caused by diffusion distorts the spatial structure, which is developed only if diffusion is slower than wave formation speed. For a fixed value of $\beta$, wave formation and propagation speed decrease as $\tau_{I}$ increases. As a consequence, the system approaches the mean-field behaviour when $\tau_{I}$ increases, as patterns are suppressed. For higher values of $\bi$, it is more difficult to sufficiently distort the patterns, such that the value of the critical point $\tic$ increases. Higher values of $D$ lower the value of the critical point: for hosts with sufficiently high local mobility, pattern formation is not possible and all the phase space is in the mean-field regime.

\begin{figure}[h]
\begin{center}
\includegraphics[width=1.\textwidth]{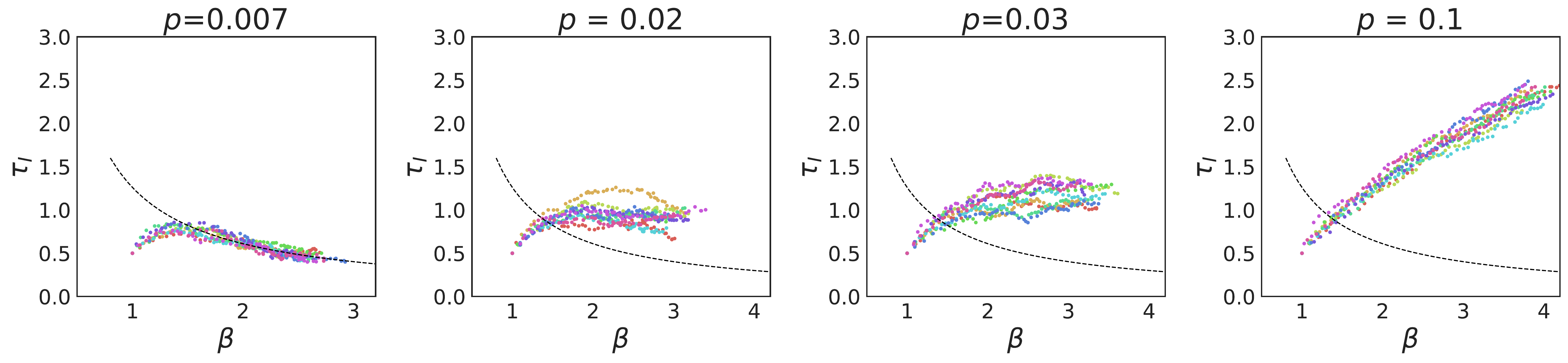} 
\end{center}
\caption{\label{fig:long_range} Evolutionary trajectories as a function of the probability $p$ of a long-range interaction. Initial conditions are $\beta_0=1.0$, $\tau_{I0}=0.5$ and $L=100$. As in the diffusive case, depending on the value of $p$ there is a transition from the mean-field phase to the ordered phase, albeit with a different value of $R_0^{ev}$. The exact value of $p$ at which the transition occurs is also dependent on the initial conditions.}
\end{figure}

The situation is qualitatively analogous if, instead of local diffusion, we assume that nodes can have arbitrary neighbours in the lattice --rather than just nearest neighbours-- with some given probability. Actually, the effect of long-distance transmission was already discussed in~\cite{ballegooijen_emergent_2004}, where it was pointed out that mixing up to $2\%$ of contacts yielded similar results, i.e. spatial structuring and an emergent trade-off with a slightly different value for $R_0^{ev}$. Our simulations indicate that, as in the case with diffusion, relatively low values of the fraction of long-distance contacts (below $10\%$ in all considered cases) prevent the formation of waves; the precise value causing the transition depends on the initial parameters. Figure \ref{fig:long_range} shows, for a fixed initial condition, the transition between the $D=0$ and the mean field behaviour, which is in fact analogous to the diffusive case.

\subsubsection{Waves develop in a finite range of $\tau_R$ values}

The intrinsic self-excitability of Ballegoijeen and Boerlijst's model~\cite{ballegooijen_emergent_2004} lies at the origin of the emerging spatial waves. This property is lost in simpler susceptible-infected (SI) models, which do not consider a recovery period after infection. In other words, spatial waves do not develop in the limit $\tau_R \to 0$. From a geometrical perspective, $\tau_R>0$ has the effect of separating subsequent wave fronts, thus allowing the formation of well-defined infection waves.

Interestingly, in the simulations, changes in $\tau_R$ do not seem to affect the critical value $R_0^{ev}$. Nevertheless, the analytic reasoning made before suggests that the critical value should change, given the dependence of equations~(\ref{eq:w}) and~(\ref{eq:expansion}) with $\tau_R$. We believe that the one dimensional argument we have used is not able to capture the full phenomenology observed in two dimensions. 

Even when $\tau_R$ does not affect the critical point of the transition, it is relevant for the study of the stability of the patterns. Our computational results show that if $\tau_R$ decreases, the transient time needed to converge to the curve $R_0=R_0^{ev}$ increases. This is due to the fact that the formation of wave fronts requires larger values of $\tau_I$ the smaller $\tau_R$ is. Therefore, for small $\tau_R$, increasingly larger values of $R_0$ are required in order to develop patterns and start the selection for wave frequency. Conversely, larger values of $\tau_R$ permit the formation of waves with lower values of $\tau_I$, thus accelerating convergence to the critical boundary. 

We have numerically studied the limit $\tau_R \rightarrow 0$, which corresponds to the conversion of our model to an SI-like model. For low values of $\tau_R$, we have checked that the system is not able to develop patterns, exhibing the mean-field behaviour as described in the diffusion case. Moreover, we have checked that too large values of $\tau_R$ also prevent the formation of waves. Intuitively, this is due to the blocking effect caused by individuals remaining for too long in the recovered state, and thus avoiding wave propagation. We cannot discard, however, that this is a finite size effect showing up when $\tau_R \simeq L$. 

In summary, considering that self-structuring patterns are possible only for intermediate values of $\tau_R$, too large or too small values of the recovery rate also jeopardize the stability of the emergent trade-off.

\subsubsection{Evolutionary trajectories are weakly affected by changes in the mutation rate}

Pathogenic organisms display a range of mutation rates that spans at least seven orders of magnitude, from the barely $10^{-10}$ substitutions per nucleotide and replication cycle of pathogenic fungi to the $10^{-4}$ of most RNA viruses~\cite{sniegowski:2000}. Viroids, circular non-coding RNA molecules that are pathogenic to plants, affecting especially crops, might present mutation rates up to over $10^{-3}$ substitutions per genome copied, thanks to the small size of their genomes (a few hundred nucleotides)~\cite{gago:2009}.

Low mutation rates are not an issue in the context of the model here studied, since they cannot disrupt the emergence of the trade-off. In a more realistic scenario, though, low $\mu$ entails long transients that would delay the convergence to the critical curve. The pertinent question, therefore, is if large values of $\mu$ prevent in any way the development of spatial self-structuring. Actually, our simulations where performed for $\mu=0.01$, which is not a small value for a phenotypic mutation rate, as implemented in the model. We have verified that mutation rates up to $\mu=0.5$ do not prevent convergence to $R_0^{ev}$, though they increase the size of fluctuations away from the critical curve. Additionally, one could consider changes in the parameters that, in agreement with empirical observations~\cite{eyre-walker:2007}, would be randomly drawn from a fat-tailed probability distribution, that is where mutations causing large changes in phenotype are not exceedingly rare. Though this possibility has not been tested, we do not expect it to modify the evolutionary trajectories in the light of our previous results. 

\subsection{Role of a metapopulation structure in the evolutionary fate of populations}

\begin{figure}[h]
\begin{center}
\includegraphics[width=0.8\textwidth]{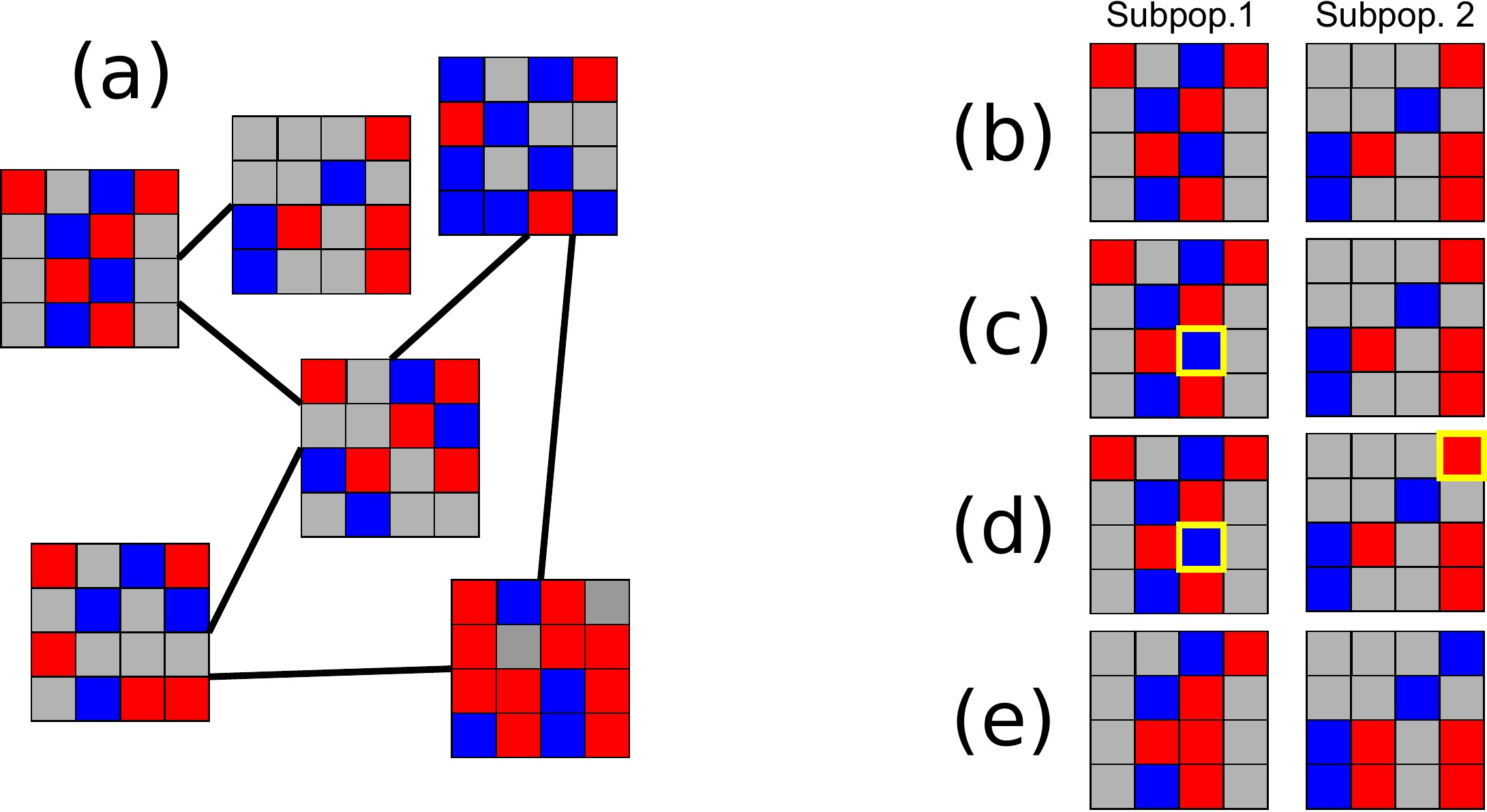} 
\end{center}
\caption{\label{fig:metapop} Metapopulation structure and site jump dynamics. (a) Example of the metapopulation structure. It consists of 25 lattices, each one defined by a diffusion $D$. We use a fully connected network for the simulations. (b) Example of two subpopulations at a given time. Each node has a probability $\lambda\Delta t/N$ of jumping. (c) A site (highlighted in yellow) in the first subpopulation is randomly selected according to the previous probability to jump to the second subpopulation. (d) For each node that jumps, a target node in the second subpopulation is selected. (e) The state of the two nodes is swapped.}
\end{figure}

Our metapopulation model consists in a network of lattices that can exchange the states of a randomly chosen pair of individuals (see Methods and Figure~\ref{fig:metapop}). Alternatively, we have also simulated a situation where infection could be propagated through vectors, and where an infected individual tries to transmit the disease to a second, randomly chosen individual in a different lattice. Since results are indistinguishable in these two formulations, here we present results for the first case (swapping of individual states). Simulations are performed with a fully-connected metapopulation (as in Figure~\ref{fig:metapop}a).

\subsubsection{The metapopulation structure is irrelevant for sessile hosts}

\begin{figure}[h]
\begin{center}
\includegraphics[width=1.\textwidth]{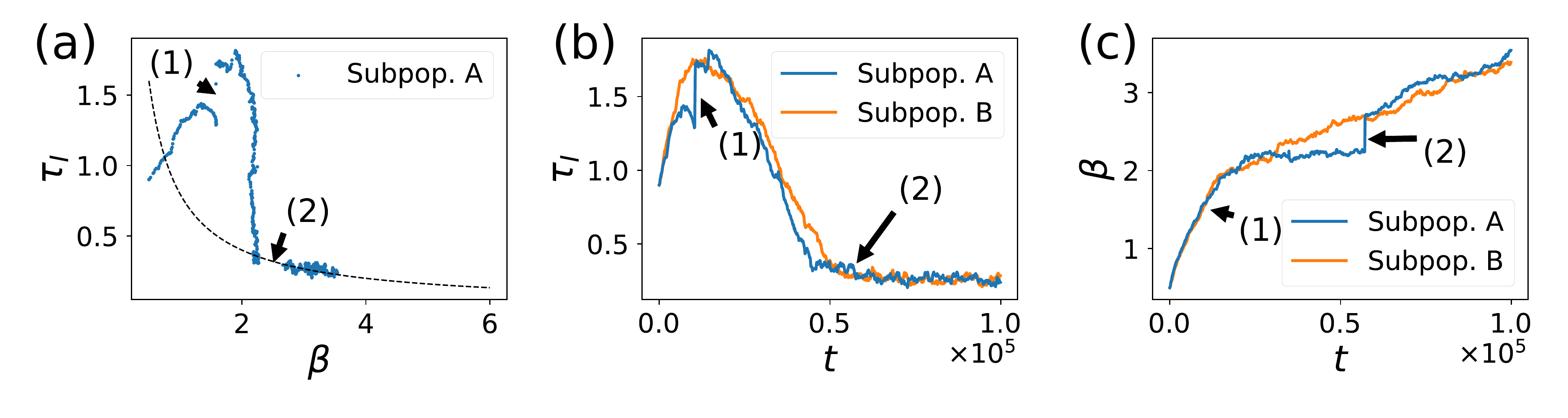}
\end{center}
\caption{\label{fig:metapop_results} Interaction between two subpopulations with diffusing hosts. We depict the detailed evolutionary trajectory of two connected subpopulations with finite diffusion $D=0.35$ in the $\beta-\tau_I$ plane (a), as well as the temporal variation of $\tau_I$ (b) and $\beta$ (c). Labels (1) and (2) correspond to events where subpopulation A was invaded by a strain of subpopulation B. In (1), the subpopulation was invaded by a strain of higher $R_0$, which was possible because pattern formation had not yet started. In (2), spatial patterns are developed and the subpopulation can only be invaded by a strain with higher frequency of emission, causing a jump forward along the critical boundary.}
\end{figure}

We performed simulations of a metapopulation where all the patches were in the no-diffusion phase or in the mean-field phase, respectively. The evolutionary trajectories are identical to the ones displayed by a single population in the no-diffusion regime or in the mean-field. This is because all the subpopulations are subject to the same selection mechanism. In fact, evolution happens locally when the patches are in any of these two limits. This result is easy to understand if we think of two connected populations, A and B. Let us say A is in the high diffusion regime and B in the $D=0$ regime. Since the base reproductive number in B has a low value, a strain that jumps from B to A will not invade the well-mixed population, where the mechanism of selection relies on $R_0$. Moreover, a strain from A to B would not invade either, since it has a low wave emitting frequency, which is the main selection mechanism in B. At the end, the evolutionary trajectory in each population is the same as the isolated system's trajectory, since the dynamics only depends on the base reproductive number of the perturbation and the spatial structure, and not on where this strain comes from. Both populations will evolve with no apparent interaction between them. Therefore, evolutionary dynamics depends only on the local spatial structure of the populations, and not on the large scale metapopulation. Figure~\ref{fig:metapop_results} illustrates this point.

\subsubsection{Well-mixed states are shortly lived in small lattices}

We have seen that, when hosts are mobile, the evolutionary dynamics does not always converge to the critical boundary but, depending on the initial conditions and the strength of diffusion, it might enter the runaway regime where $R_0$ steadily increases. The eventual fate of such populations is a pathogen-free system with all individuals in the susceptible state. While in the absence of other populations that may act as reservoirs of the pathogen this is the final state, pathogen-free populations can be rescued if a metapopulation structure is present. At this point, therefore, the fact that populations in the mean-field regime have a finite lifetime becomes an important evolutionary feature. This is in contrast with self-structured populations, which are in a state of endemic infection that is sustained in time. 

\begin{figure}[h]
\begin{center}
\includegraphics[width=0.6\textwidth]{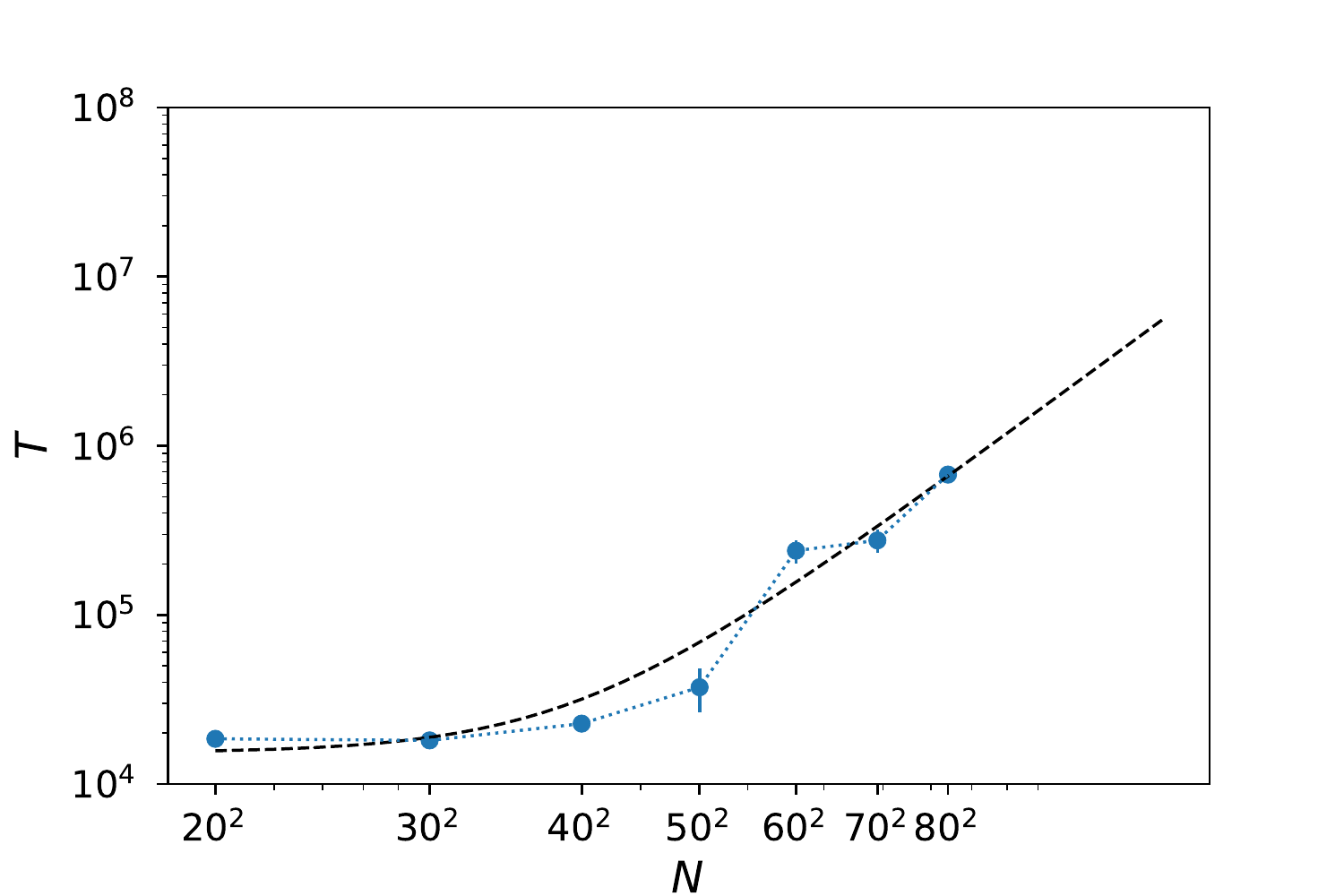} 
\end{center}
\caption{\label{fig:scaling}
Lifetime of populations in the mean-field phase. The diffusion is $D=0.35$, and lifetime is represented as a function of the system size $L$ in log-log scale. Points are averages over 10 realizations. The best fit to data is shown with a dashed line, $T(N)=(5.3\pm3.0)\cdot 10^{-5}N^{(2.6\pm0.6)} + (1.5\pm0.3)\cdot 10^4$ as obtained through least-squares fit.}
\end{figure}

With this motivation in mind, we have studied the average lifetime of populations for intermediate values of the diffusion as a function of their lattice size. Here, lifetime is defined as the number of timesteps required for all the individuals to reach the susceptible state. As expected, populations in the ordered phase did not decay in any of the performed simulations. In contrast, the lifetime of populations as a function of their lattice size can be fitted in the mean-field phase to a power-law with exponent $(2.6\pm0.6)$ (Figure~\ref{fig:scaling}). In the limit of infinitely large systems, the mean-field phase is stable and infection can survive for very large times whose duration diverges as $L \to \infty$.

\subsubsection{The metapopulation structure favours the emergence of the evolutionary trade-off if hosts diffuse}

The results in the previous section quantify differences in the lifetime of subpopulations that have reached the critical boundary, which are self-structured and thus stable in time, or entered the mean-field behaviour, therefore with a finite lifetime that depends on their size. At some point, the latter will become fully susceptible and can be re-infected by a neighbouring population. When hosts are mobile, the reinfected lattice can either be attracted again to the mean-field behaviour or develop waves and converge to the critical boundary. The likelihood of either outcome is dependent on the initial conditions, as we have shown. However, the initial conditions are not arbitrary now, but correspond to the state of a neighbouring lattice, which has already evolved to one of the two possible states. If the lattice from which the infecting individual is drawn is the mean-field regime, the newly infected lattice will also fall into that disordered phase, and again collapse in finite time to the fully susceptible situation. Instead, if the infecting individual is drawn from a subpopulation that has converged to the critical value $R_0^{ev}$, the initial conditions are such that the newly infected lattice will develop spatial patterns and the emergent trade-off. The ability of a neighbouring individual to infect follows the criteria derived for mutants in former sections (Figure~\ref{fig:metapop_results}). 

In a metapopulation structure, however, the global process functions as a ratchet. Once a lattice has settled in a self-structured state with fixed base reproductive number $R_0^{ev}$, it will not be kicked out of it by any attempt of invasion from neighbouring subpopulations. At the same time, populations in the disordered state regularly collapse until they are infected in such conditions that they fall into the ordered phase, at which point their dynamics are stable and long-lived. Therefore, a metapopulation structure selects for a globally ordered phase due to the finite lifetime of any subpopulation in the disordered phase, where $R_0$ diverges. 

\section{Discussion}

In this work, we have revisited a spatial self-structuring model of host-pathogen evolution where an emergent trade-off between infectivity and transmissibility had been described~\cite{ballegooijen_emergent_2004}. We have shown that an evolutionary feedback mechanism here characterized poises the system to a critical boundary where selection on the base reproductive number and selection on the frequency of emission of infective waves equilibrate. This critical, self-evolved state only emerges if the conditions of the system are such that spatial ordering in the form of epidemic waves can set in. The critical boundary separates two regions characterized by spatial order/disorder through a phase transition where incipient spatial waves wax as disorder wanes. The precise, quantitative nature of the phase transition, however, needs to be formally characterized in future research. 

Classical definitions of fitness often fail in spatially extended evolutionary competition, as the inability of $R_0$ or $w$ to predict by themselves the winner in an invasion here demonstrates. Indeed, there have been other studies clearly pointing out that, in spatially structured evolving populations, the strategy of maximizing $R_0$ is not adaptive in the long run, either because fitness depends on the time-scale (different strategies are successful at different times)~\cite{rauch:2002} or because, at odds with mean-field scenarios, a finite fraction of immune hosts might induce pathogen extinction even if $R_0$ is arbitrarily large~\cite{cuesta:2011}. 

Convergence to the critical boundary is a far from trivial issue. First, the emergence of the trade-off can be severely delayed depending on initial conditions and specific evolutionary parameters. Second, the trade-off is structurally unstable under host mobility, such that moderate values of host difusion or of a fraction of long-distance host jumps cause a cross-over to a mean-field behaviour, where the base reproductive number grows unboundedly. Third, the evolutionary stable, finite $R_0^{ev}$ value can be achieved through successive invasions of the resident pathogen only if invasion is attempted at a local scale. This result is highly reminiscent of invasion experiments by mutant viral strains where it was observed that the substitution of the wild type by an in principle fitter mutant did not succeed unless the mutant was seeded above a minimum relative population size threshold~\cite{delaTorre:1990}. In the latter case, the impossibility to displace the wild type if the invading population was too small was ascribed to its limited genotypic heterogeneity~\cite{aguirre:2007}. Actually, space might play an important role in the emergence of heterogeneous populations and its properties~\cite{aguirre:2008}, therefore conditioning also in this respect evolution~\cite{rauch:2003} and invasion~\cite{champagnat:2007}. Fourth, convergence to the critical boundary is dependent on the initial evolutionary parameters, a fact that might have important ecological implications. It indicates that the onset of spatial self-structuring, and therefore of the emergent trade-off, is contingent on the life-history of the system, and afects its evolutionary fate. Indeed, the jump of a pathogenic species to a new host is affected by multiple variables, among which ecological factors, viral genetic plasticity and host specificities~\cite{elena:2011}. While sufficiently long coevolution with the original host species has probably selected for evolutionary parameters permitting coexistence, the aetiology of the disease might be completely different in the new host. If the pathogen turns out to be too virulent (i.e., it starts with a too high $\tau_I$), persistent infection of this new host is prevented, and it can only infect in bursts that terminate with the death of the local host population and the erradication of the pathogen in a short time. Bursts of infection can however have different origins, and in particular result from prudent infective strategies~\cite{boerlijst_spatial_2010}.

In a metapopulation organization, we have shown that the evolutionary dynamics happen at the scale of the local populations, and not at the scale of metapopulations, in contrast with other studies where the dynamics of the metapopulation cannot be extrapolated from the dynamics of a patch (see e.g.~\cite{poletto_characterising_2015,poletto_host_2013}). Moreover, in the current case evolutionary trajectories in patches can be unrelated if the subpopulation properties are very different, leading to a sort of ecological speciation. Instead, the metapopulation structure is here responsible for driving the system to a globally stable phase in the presence of host mobility. In this study, we have kept the jump rate between subpopulations fixed for all strains, though, together with a complex network structure, it may affect the spreading of disease in the metapopulation at long time scales.

Though attention is typically focused on the evolution of pathogenic traits and on the immune strategies of the host (be they intrinsic, through an immune coevolving system or extrinsic, as in avoiding behaviour), it cannot be discarded that the spatial pattern itself be a feature under selection~\cite{jackson:2014}. In a different class of systems, it has been shown that disordered states are conductive to extinction, as in the case of spatially extended catalytic hypercycles, where the formation of spiral waves avoids the otherwise lethal effect of parasitic mutants~\cite{boerlijst:1991}. The question therefore remains, whether selection for long-lived coexistence with the host mediated by the selection of specific spatial patterns may act as an additional force to promote emergent trade-offs.

\section*{Methods}
Here we describe in detail the numerical algorithm used to simulate the dynamics of the model, including the diffusion scheme. The steps of the algorithm are:

\begin{enumerate}
\item Initialize the lattice with periodic boundary conditions, and a $5\%$ of infected individuals, all of them with the same initial strain. Take $t=0$. Initialize internal counters $t_{j}=0$ for all nodes. 
\item At each timestep, we iterate over the nodes. For node $j$, 
\begin{enumerate}
\item If the node is susceptible, consider the set of infected neighbours $\Omega_{j}$. Node $j$ is infected with probability $p=1-\exp\left(-\Delta t\sum_{n\in\Omega_{j}}\beta_{n}\right)$. Each neighbour has different transmission rate, so we have to select the origin of the infection computing all the probabilities $p_{m}=1-\exp\left(-\Delta t\beta_{m}\right)$ and then selecting the source node with probability $q_{m}=p_{m}/\sum_{n\in\Omega_{j}}p_{n}$ to ensure normalization. If neighbour $m$ was selected, then set $\beta_{j}=\beta_{m}$ and $\tau_{Ij}=\tau_{Im}$. Using this method, we infect the node at the correct rate, and select the neighbour with a probability according to its transmission rate. 
\item If the node is infected, check if $t_{j}\geq\tau_{Ij}$. In this case, the infection has finished, the node becomes recovered, and internal time is reset to $t_{j}=0$. In other case, with probability $\mu\Delta t$ make a mutation $(\beta_{j},\tau_{Ij})\rightarrow(\beta_{j}\pm\Delta\beta,\tau_{Ij}\pm\Delta\tau)$, changing $\beta_{j}$ and $\tau_{Ij}$ independently. 
\item If the node is recovered, check if $t_{j}\geq\tau_{R}$. If this happens, then make the node susceptible again. 
\end{enumerate}
\item We have $t=k\Delta t$. If $k=\Gamma$ then apply the mixing algorithm. 
\item Make $t=t+\Delta t\equiv(k+1)\Delta t$ and return to step 2 until
the desired time. 
\end{enumerate}
The next step is to select the parameters. We choose to fix $\tau_{R}=1$ as  the basic timescale (except when this parameter is externally varied). All times are relative to this scale. We have fixed $\Delta t=0.01$, $\mu=0.01$, $\Delta\beta=0.01$ and $\Delta\tau=0.01$, as in~\cite{ballegooijen_emergent_2004}. Lattice size varies through the study and is indicated in each case.



\bibliographystyle{naturemag} 
\bibliography{refs}

\newpage
\appendix

\section*{Finite size effects}

In our experiments with diffusion, we can observe the ordered and the disordered phases coexisting in a finite interval around $\tic$. This is due to the existence of stochastic fluctuations that arise for a finite system size $L$. We have checked that increasing the system size reduces the width of the interval, so in the thermodynamic limit $L\rightarrow+\infty$ we only have a point separating the two phases. 

A different effect of finite systems regards the appearance of periodic patterns which we here describe. Their appearance is due to the use of periodic boundary conditions, which permit that the characteristic length of patterns couples with system size $L$. In our experiments, we have taken $L$ large enough so finite size effects can be neglected. 

\begin{figure}[h]
\begin{center}
\includegraphics[width=1.\textwidth]{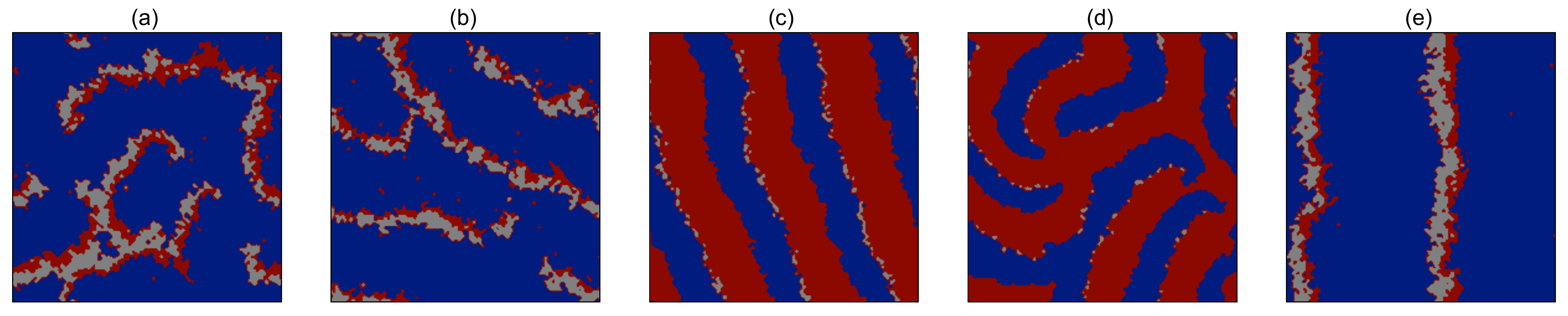}
\end{center}
\caption{\label{fig:finitesize}
Finite size effects. We show the system configuration for large times, when finite size effects happen, for a system size $L=100$. Time advances from left to right. (a) Spiral wave patterns before finite size effects start. (b) Spiral waves start to couple with system size, reaching both sides of space. (c) Wide linear fronts. (d) Collapse of the linear fronts into spirals. (e) Stable linear fronts.}
\end{figure}

When we let the system evolve without diffusion, strains that emit infected waves with higher frequency are selected. As the frequency changes, it becomes more probable that a wave can "connect" two sides of the space, producing a linear front. Inside the linear front, strains tends to increase its infection time, leading to the formation of stripes. Eventually, the high $R_0$ achieved at some points in the stripe deforms it and breaks the wide fronts again. Some pictures of the process are shown in Figure~\ref{fig:finitesize}.

To understand better this phenomenon, we have looked for a characteristic length of the system. This is done for several times $t$ as follows:

\begin{enumerate}
\item{Assign numerical values to the states, susceptible $\rightarrow 0$, infected $\rightarrow 1$, recovered \textbf{•}$\rightarrow -1$ so we can operate with them. We have checked that the particular selection of values does not affect the results qualitatively.}
\item{Compute the two-dimensional Fourier transform of the system.}
\item{Compute the most important Fourier modes, $|k|$, and define a characteristic length as $\lambda = 1/|k|$.}
\end{enumerate}

If the evolutionary mechanism is not active, the characteristic length is always constant and depends on the emission frequency, meaning that $\lambda$ can account for the evolution of the observed patterns. For the evolutionary trajectories, we have found that the characteristic length tends to increase with time, up to the point where the fronts form. After this point the characteristic length becomes constant. 
We have also checked that the time needed to start producing the fronts increases as the system size $N$ is increased. Therefore, our conclusion is that the front formation is a finite-size effect due to a coupling between a characteristic scale of the system and the system size. For this reason, in the experiments system size and timescales have been selected such that no finite-size effects can be seen. This effect was avoided in the original paper by Ballegooijen and Boerlijst [{\it Proc. Natl. Acad. Sci. USA} 101(52)18246--18250 (2004)] thanks to open boundary conditions. The nature of the boundary, therefore, does not play any role in the phenomenology discussed.

\newpage

\section*{Figure caption}

\end{document}